\documentstyle[epsf,epsfig,rotate]{aa}
 
\newcommand{\be}{\begin{equation}}
\newcommand{\ee}{\end{equation}}

\newcommand{\ls}{\; \raisebox{-.8ex}{$\buildrel{\textstyle<}\over\sim$}\;}
   
\newcommand{\anrev}{{\it ARA\&A, }}

\newcommand{\mnr}{{\it MNRAS, }}

\newcommand{\bmth}[1]{\mbox{\boldmath${#1}$}}

\newcounter{pp3}
\addtocounter{pp3}{3}
   
\begin{document}
     
\title
{ The Local Instability of Steady  Astrophysical Flows
with non Circular Streamlines with Application to 
Differentially Rotating Disks with Free Eccentricity}

\author{J.C.B. Papaloizou}
\institute{ Astronomy Unit,
Queen Mary, University of London, Mile End
Rd, London E1 4NS}
	
\offprints{jcbp@maths.qmw.ac.uk}

\date{Received /Accepted}
   
\def\LaTeX{L\kern-.36em\raise.3ex\hbox{a}\kern-.15em
T\kern-.1667em\lower.7ex\hbox{E}\kern-.125emX}
\titlerunning{   Eccentric Disks}
\authorrunning{ J.C.B.   Papaloizou }

\abstract{
We carry out a general  study  of the
stability 
of astrophysical flows that appear steady
in a uniformly rotating frame.
Such a flow might correspond to a stellar pulsation  mode
or an accretion disk with a free global distortion
giving it finite eccentricity . 

\noindent We consider perturbations arbitrarily 
localized in the neighbourhood of unperturbed fluid streamlines.
When conditions do not vary around them, perturbations  take the form of 
oscillatory inertial or gravity modes. However, when conditions do vary
so that a circulating fluid element is subject to periodic variations,
parametric instability may occur. For nearly circular streamlines,
the dense spectra associated with inertial or gravity  modes 
ensure  that resonance conditions can always
be satisfied when twice the period of circulation round
a streamline
falls within.  

\noindent We apply our formalism to a  differentially rotating disk for which
the streamlines are Keplerian ellipses, 
with free eccentricity up to 0.7, which do not precess in an inertial frame.
We show that
for small $e,$
the instability 
involves parametric excitation of
two modes with azimuthal mode number
differing by unity in magnitude which have a period
of twice the  period of variation  as viewed from a circulating unperturbed  fluid element.
Instability  persists over a widening range
of wave numbers with increasing growth rates  for larger eccentricities. 

\noindent The nonlinear outcome is studied in a follow up paper which
indicates development of small scale subsonic turbulence.

\vspace*{5cm}
}
\maketitle

\section {Introduction}

\noindent  In Astrophysics one is often concerned with the stability
of a system undergoing some regular global steady motion. Possible 
 examples are stars undergoing a pulsation that may be either freely excited
or forced through tidal interaction in a binary system.
 Other examples, which are more the focus of this paper,  are differentially rotating disks
which have been either set up or perturbed into a state in which the orbital
motion about the centre is non circular. Systems of this type are known to
be susceptible to parametric instabilities which involve a resonance 
between the periods of normal modes of  oscillation  and  a period
associated with the unperturbed motion.   For examples
of studies related to stellar oscillations 
see Vandakurov(1967), Papaloizou (1973), Papaloizou \& Pringle 1980),
Kumar \& Goodman (1996), and Wu \& Goldreich (2001).

\noindent The presence of accretion disks in which the orbits are eccentric have been
suggested in a number of contexts. The production of such disks by the tidal
disruption of stars orbiting around AGN has been considered by 
Gurzadyan \& Ozernoy (1979). A model to explain the long term  spectral variability
of Be stars as being due to global one-armed $(m=1)$ oscillations of equatorial disks
has been developed  (eg. Okazaki 1991; Papaloizou, Savonije  \& Henrichs 1992).
Eccentric disks have also been proposed as an explanation for the superhump
phenomenon in SU UMA stars (Whitehurst 1988).

\noindent A general theory to describe the slow viscous evolution of disks with modest
to large eccentricities has been developed by Ogilvie (2001). This confirms the possibility
of long lived structures of this type within that context.

\noindent The stability of the non circular motions induced in an accretion
disk by a binary companion has been studied in pioneering work by
Goodman (1993) and Ryu \& Goodman (1994). These authors
studied the fast $m=2$ forced distortions that propagate around with the companion.
A parametric instability was found  that lead to turbulence
in the context of local  two dimensional
modelling in a shearing box.

\noindent  In this paper we are concerned with extending studies
of the stability of disks with non circular motions to more general situations
 such as  global disks with free
global $m=1$ modes which are almost stationary in an inertial frame.

\noindent The recent discovery of a number of extrasolar giant planets orbiting around
nearby solar--type stars ( Mayor \& Queloz 1995;
Marcy \& Butler 1998, 2000) 
with high orbital
eccentricities in the range $ 0.0 \; \ls \; e \; \ls \; 0.93$
has led to the consideration of the  proposal 
that a giant planet orbiting  in and interacting with  a protostellar
disk (eg. Goldreich \& Tremaine 1980;  Ward 1997) 
 could undergo an instability that would cause its eccentricity to increase.
(Artymowicz 1992; Lin \& Papaloizou 1993; 
 Papaloizou  Nelson \& Masset 2001;  Goldreich \& Sari 2003).
However, the free modes in such an interaction are likely to 
be coupled to the disk which participates with significant eccentricity
in a global $m=1$ mode.
(Papaloizou 2002, Goldreich \& Sari 2003). Accordingly the behaviour
and in particular the damping of free  eccentricity in differentially rotating disks
becomes an issue. It is  one of the purposes of this
and  a follow up paper (Papaloizou 2004)  to study the instabilities
of free global  disk normal modes corresponding to free eccentricity
and their non linear outcome. Although there may be major  regions of protostellar
disks that support magnetic fields leading to turbulence through the Magnetorotational
instability (Balbus \& Hawley 1991), because of dependence on the existence of  external sources of ionization
 and complex chemistry,  the extent of such regions is highly 
uncertain ( eg. Gammie 1996 ; Fromang, Terquem \& Balbus 2002).
Accordingly for this  first study we have considered disks with eccentricity
in the hydrodynamic limit. This is of general theoretical interest
because even in that limit there are potential local  instabilities of disks with non
circular but regular streamlines that could lead to turbulence and enhanced dissipation
(Goodman 1993). Although uncertain, it is likely that the presence of magnetic fields would 
enhance any decay rate found in the hydrodynamic limit.  However, this will be 
studied in a separate investigation.

\noindent In this paper we develop a general  formalism for the study
 of the local 
stability
of astrophysical flows possibly in stars or disks that appear steady
in a uniformly rotating frame.

\noindent We adopt a local Lagrangian approach and  consider perturbations
localized  near unperturbed fluid streamlines.
These are shown to be governed by a pair of second order ordinary differential
equations that give the evolution of the localized disturbance in the
neighbourhood of a fluid element as a function of time. 
 For streamlines on which conditions vary as a fluid element 
circulates, the coefficients are periodic functions of time.
The parametric excitation of inertial or gravity modes readily occurs.

\noindent We solve these equations and delineate the 
instability for an example
for which the streamlines are Keplerian ellipses,
with free eccentricity up to 0.7. 
\noindent The nonlinear outcome of such instabilities 
in disks supporting global modes which endow
them with eccentricity
is studied in a follow up paper which
indicates development of small scale subsonic turbulence.

\noindent The plan of the paper is as follows:
In section \ref{s1} we give the Basic equations.
In section \ref{Stmodel} we describe the steady state flow conditions
of the systems  whose stability we analyse, in the context
of a model with streamlines that are Keplerian ellipses.
In section \ref{Linstab} we discuss the general linear  stability analysis
and go on to describe the solution procedure used in the local limit
in  general  terms relating the spectra found in the local
problem to that of the global one in  appendix 3.
In section \ref{sime} this is applied to 
the case  when there are circular streamlines in the
unperturbed  steady state. 
In section \ref{notsim} this is extended
to a steady state model with streamlines taking the form of Keplerian
ellipses.
Numerical and analytic solution of the governing equations
are described in section \ref{govsol} for the Keplerian streamline case
while emphasizing the general nature of the behaviour. 
Finally in section \ref{Discu} we discuss our results

\section{Basic Equations}  \label{s1}
We work in a  rotating  frame which   rotates with 
angular velocity $\bmth{\Omega_P} $  relative
to an inertial frame.

\noindent We adopt a system 
of cylindrical coordinates $(r,\varphi,z)$ for which
the origin is at the centre of mass and  
the angular velocity is directed along the $z$ axis
such that $\bmth{\Omega_P}  = ( 0, 0, \Omega_P).$

\noindent In the absence of magnetic fields, the  equation of motion 
may be written in the form

\begin{equation}  
{\partial  {\bf v} \over \partial t}+ {\bf v}\cdot \nabla {\bf v}
+ 2\bmth{\Omega_P}\times {\bf v}
+\bmth{\Omega_P}\times \left (\bmth{\Omega_P}\times {\bf r}\right )
= {\bf F}= -{\nabla P \over \rho} - \nabla \Psi .
\label{1em} \end{equation}

\noindent Here the velocity is ${\bf v} = ( v_r , v_{\varphi}, v_z).$
The left hand side includes contributions from
the acceleration of a fluid element and coriolis and centrifugal
accelerations arising because of the rotating frame. 
The acceleration of a fluid element may be written
using the  convective  derivative
taken moving with the fluid in the form

\begin{equation}{D {\bf v} \over D t} =
{\partial  {\bf v} \over \partial t}+ {\bf v}\cdot \nabla {\bf v}.
\label{1conv} \end{equation}

\noindent The system is subject to a force per unit mass ${\bf F}$
which has components due to pressure, $P,$ and gravity, with 
the gravitational potential being $\Psi.$ The density is $\rho.$

\noindent In addition we have the  continuity equation
\begin{equation}
{\partial \rho \over\partial t}+
\nabla \cdot (\rho {\bf v} )
=0.\label{1cont} \end{equation}

\section{Steady State Configurations} \label{Stmodel}
We here consider the stability of steady state
configurations for which $\partial  / \partial t \equiv 0.$
For these equation (\ref{1em}) gives

\begin{equation}
{\bf v}\cdot \nabla {\bf v}
+ 2\bmth{\Omega_P}\times {\bf v}
+\bmth{\Omega_P}\times \left (\bmth{\Omega_P}\times {\bf r}\right )
= {\bf F}= -{\nabla P \over \rho} - \nabla \Psi .
\label{1sst} \end{equation}

\noindent In general this allows for a steady state flow ${\bf v}$
as viewed in a rotating frame. 

\noindent A very simple  possible motion
is non uniform rotation for which ${\bf v} = (0, v_{\varphi}(r), 0)$
corresponding to circular motion about the $z$ axis at a rate which is a function of $r.$

\noindent Other examples could correspond to a normal mode of stellar oscillation as viewed
in a frame rotating with the modal pattern speed or an accretion disk
with a global eccentricity.

\subsection{Cylindrical Disks}

\noindent  In the context of differentially rotating
accretion disks we consider
fluid circulating in Keplerian ellipses in a cylindrical
potential. For this $\bmth{\Omega_P}=0$ and  
$\Psi = -GM_*/r.$
Here $M_*$ is a central mass and $G$ 
the gravitational constant. Disk self gravity is neglected.
If in addition the pressure, $P,$ is constant, we have 
\begin{equation}
{\bf v}\cdot \nabla {\bf v} =
- \nabla \Psi .
\label{1ssimpt} \end{equation}
This model can approximate conditions near the midplane of a vertically stratified  disk
if $\partial v_z / \partial z =0$ there and with a slight modification 
if it is not.

\noindent Equation (\ref{1ssimpt}) applies to particle motion under
only the potential $\Psi.$ Thus a solution can be found
from the Keplerian trajectory 
\begin{equation}
{a(1-e^2)\over r} =1 + e\cos \varphi , \label{orbit}
\end{equation}
where the semi major axis $a$ and the eccentricity $e$ are constant
on a streamline being a particle trajectory.
On such a trajectory the velocity components are given by
\begin{equation}
{v_r } =  e\sin \varphi \sqrt{ { GM_* \over  r(1+e\cos\varphi)}} , \label{radvel}
\end{equation}
and
\begin{equation}
{v_{\varphi} } = \sqrt{GM_* (1+e\cos\varphi) \over  r}.\label{phivel}
\end{equation}

\noindent The eccentricity is constant on a stream line and may be specified
as an arbitrary  function of $a.$ Equation (\ref{orbit}) may then be solved
for $e$ as a function of $r$ and $\varphi.$ The density may then be found from
the steady state continuity equation 
$\nabla \cdot (\rho {\bf v}) = 0.$ In the special case when $e$ is constant,
the continuity equation is satisfied for $\rho$ being  a general function
of its arguments of the form

\begin{equation}
\rho  = \rho \left( r(1+e\cos\varphi), z \right).
\end{equation}
 
\noindent For the work presented later on we adopt a locally isothermal
equation of state such that 

\begin{equation} P= \rho c_s^2(r). \label{EOS}\end{equation}

\noindent The sound speed $c_s(r)$
 is taken to be a fixed function of $r.$
As above we  adopt a cylindrical disk model for which
vertical stratification is neglected
(eg. Hawley 2000; Steinacker \& Papaloizou 2002)  and 
the potential is independent of $z$ such that
$\Phi = -GM_*/r.$ We note  again that this
model can describe conditions close to the midplane of a stratified disk
which is all that is required in order to perform a local stability
analysis of the type we consider here.
When $\rho$ is constant and $c_s^2(r) \propto 1/r,$
Keplerian ellipses are possible streamlines that form a stationary
pattern in the inertial frame.

\subsection{ Properties of Streamlines}\label{stream}

We study the stability of the steady state configurations described above, below.
We focus on perturbations that are localized on particular streamlines
and we assume that the configuration of steady state streamlines
is smooth enough that at least locally the fluid volume can be filled
with stream tubes that can be used to define a volume element
$dV = d{\cal A} ds,$ with $d{\cal A}$ being the  stream tube area element normal
to  a streamline with element of length $ds.$  
Then the steady flow condition $\nabla \cdot \rho  {\bf v} =0$ on a streamline 
may be written  $d(\rho |{\bf v}|{\cal A} )/ds=0.$
We also suppose that two independent invariants can be defined
on each streamline. For such  invariants $\zeta_i, i=1,2$
\begin{equation}
{\bf v}\cdot \nabla \zeta_i = 0.
\end{equation}

\noindent In the simple case of differential rotation about the $z$ axis,
the stream lines are circles centred on the $z$ axis.
The invariants can be taken to be $r $ and $z,$ the radius of a particular circle
and its height $z$ respectively. 
Similarly for the cylindrical potential models with streamlines being
Keplerian ellipses. The invariants could be taken to be $z$ along
with the semi-major axis or eccentricity, both of which are constant on a streamline.

\section{Linear  Stability Analysis} \label{Linstab}

\subsection{General Formalism}

We now study the stability of  steady state flows of the
type introduced above. The most convenient formalism 
is the Lagrangian approach developed by Lynden-Bell \& Ostriker (1967).
Following these authors
we introduce the Lagrangian variation $\Delta $
such that  any state variable $Q$ is perturbed   following  a fluid element  from
its value appropriate to the steady state  
such that
\begin{equation}
Q \rightarrow Q + \Delta Q .
\end{equation}
The Lagrangian displacement  is introduced as the Lagrangian variation
of the position vector of a fluid element so that
\begin{equation}
\Delta {\bf r} = \bmth{\xi} \end{equation}
The Lagrangian variation of the velocity is then
\begin{equation}
\Delta {\bf v} = {D \bmth{\xi} \over D t},\end{equation}
where the convective derivative of $\bmth{\xi}$ is
\begin{equation}{D  \bmth{\xi} \over D t} =
{\partial \bmth{\xi}\over \partial t}+ {\bf v}\cdot \nabla  \bmth{\xi}.
\label{svconv} \end{equation}
and as shown by Lynden-Bell and Ostriker(1967)
\begin{equation}
\Delta \left( {D {\bf v} \over Dt} \right )
={D \Delta {\bf v} \over Dt} ,\end{equation}

\noindent The Eulerian variation, $Q'$  in $Q$  seen in a fixed coordinate
system  resulting from $\Delta Q$
is given by
\begin{equation}
Q'=  \Delta  Q -  \bmth{\xi}\cdot \nabla  Q.
\label{Eul} \end{equation}

\noindent To obtain the basic governing equations for the linear stability
problem, we take the Lagrangian variation of the
equation of motion (\ref{1em}) which gives.
\begin{equation}
{D^2\bmth{\xi} \over D t^2}
+ 2\bmth{\Omega_P}\times {D \bmth{\xi} \over D t}
+\bmth{\Omega_P}\times \left (\bmth{\Omega_P}\times  \bmth{\xi}\right )
= \Delta {\bf F}.
\label{1sem} \end{equation}
Using (\ref{Eul}) we obtain
\begin{equation}
\Delta {\bf F}= -{\nabla P' \over \rho} + {\rho' \over \rho^2}\nabla P 
-\bmth{\xi}\cdot\nabla\left({\nabla P \over \rho} + \nabla \Psi\right).
\label{Feq}\end{equation}
We comment that because of the local nature of the perturbations
considered, we use the Cowling
approximation in which  variation of the gravitational potential is neglected.

\noindent The Lagrangian variation in the density is related to the 
Lagrangian displacement through

\begin{equation}
\Delta \rho = \rho' +\bmth{\xi}\cdot\nabla\rho =  - \rho\nabla\cdot\bmth{\xi} .
\end{equation}

\noindent The pressure and density variations could be related by the adiabatic
condition
\begin{equation}
{\Delta P \over P} = {\Gamma \Delta \rho \over \rho} ,\label{adiab}
\end{equation}
where $\Gamma$ is the first adiabatic exponent. For simplicity
we shall assume this to be constant below, however this is
not an essential requirement for what follows.

\noindent One can seek solutions for which the time dependence is through
a factor $\exp(i\sigma t)$ where $\sigma$ is the eigenfrequency
associated with a normal mode and after taking all time derivatives
divide this factor out. Then
\begin{equation} {D \bmth{\xi} \over D t} \equiv 
i\sigma \bmth{\xi} + {\bf v}\cdot \nabla  \bmth{\xi}.
\label{ssvconv} \end{equation}
and
\begin{equation} {D^2 \bmth{\xi} \over D t^2}  \equiv
\left( -\sigma^2 + 2i\sigma {\bf v}\cdot \nabla \bmth{\xi}
+{\bf v}\cdot \nabla\left( {\bf v}\cdot\nabla\bmth{\xi} \right) \right).
\label{ssvconv2} \end{equation}

\noindent Adopting equations (\ref{ssvconv} - \ref{ssvconv2}), the linear
stability problem (\ref{1sem}) can be reduced to the operator equation
(see Lynden-Bell \& Ostriker 1967)

\begin{equation}
{\bf L}(\bmth{\xi}) = -\sigma^2\bmth{\xi} +\sigma {\bf B}(\bmth{\xi}) 
+ {\bf C}(\bmth{\xi}) = 0, \label{LBO}
\end{equation}
where, if the adiabatic condition (\ref{adiab}) 
is used, the operators ${\bf B}, {\bf C}$
are self adjoint with weight $\rho.$ 
That is for two arbitrary  displacements $(\bmth{\xi}, \bmth{\eta})$
and standard boundary conditions

\begin{equation}
(\bmth{\eta}, {\bf C}( \bmth{\xi) }) \equiv 
\int_V \rho \bmth{\eta^*}\cdot {\bf C}(\bmth{\xi})dV =
\int_V \rho \bmth{\xi}\cdot {\bf C^*}(\bmth{\eta^*})dV,\label{iprod} 
\end{equation}
with the integration being taken over the domain 
of the fluid and a similar relation holding for ${\bf B}.$
This also defines the inner product when ${\bf C} = {\bf I},$
the identity.

\noindent This self adjoint property, although useful for some purposes
is not essential to the analysis carried out in this paper
which can in fact be extended to equations of state for which
it does not hold (see below).
Accordingly  we shall not consider the self-adjoint 
formalism further here.

\subsection {Local Analysis}\label{loco}
We here consider perturbations that are localized on streamlines.
To do this we suppose that any perturbation quantity 
is of the form 
\begin{equation}
\Delta Q = (\Delta Q)_0 W(\zeta_1,\zeta_2)\exp(\i\lambda S(\zeta_1, \zeta_2)).
\label{locco}
\end{equation}
Here the phase function $S$ and the positive definite
localization function $W$ are functions of the streamline
invariants $\zeta_1$ and $\zeta_2.$ The constant $\lambda$ is a large parameter.
The effective wave number is 
\begin{equation}
{\bf k} = \lambda \nabla S.
\end{equation}
On a particular streamline with $\zeta_1 = \zeta_{10}$ and
$\zeta_2 =\zeta_{20}$ say, this is a general linear
combination of vectors in the $\nabla \zeta_1$ and $\nabla \zeta_2$ directions.
These are independent because $\zeta_1$ and $\zeta_2$ are.
The parameter $\lambda$ can be used to make the length scale
$1/|{\bf k}| \propto 1/\lambda $ arbitrarily small.
The localization function $W$ is chosen to approach zero as one moves away from a particular
streamline  on a length scale that approaches zero as $\lambda \rightarrow \infty$
but more slowly than $1/|{\bf k}|$ say  as $\propto 1/\lambda^{1/2}.$  

\noindent The amplitude factor $(\Delta Q)_0$ is assumed to be a
function of time and  slowly varying in space.
Because of the localization this is effectively along the streamline
with $\zeta_1 = \zeta_{10}$ and
$\zeta_2 =\zeta_{20}.$

\noindent Adopting the above form of perturbations,
 because of the rapid variation
of the complex phase $\lambda S$, one can make the standard assumption of local analysis
that only the variation of this needs to be considered when taking
spatial  derivatives.
The only exception occurs when the variation is only along
streamlines. This happens when the  derivatives in an expression
occur in the form ${\bf v} \cdot \nabla$
as in for example equations(\ref{ssvconv} - \ref{ssvconv2}).
Then, because ${\bf v} \cdot \nabla S =0,$ 
the contribution ${\bf v} \cdot \nabla (\Delta Q_0)$ must be retained.

\noindent Proceeding in this way one may obtain a set of governing
equations for quantities on a streamline on which the perturbation is localized.
Because the exponential and localization factors cancel out
we drop the subscript zero from the perturbation amplitudes.
Apart from such a justification  as in standard local analysis one can also prove that
the eigenvalues $\sigma$ one obtains belong to the spectrum of the  global
linear stability problem
(see appendix 3 
and also Papaloizou \& Pringle 1982; Terquem \& Papaloizou 1996 for a
discussion of related problems).
\subsection{Reduction of the Linearized Equations}
Before embarking on a local analysis, we perform a prior reduction
of the stability problem to a convenient form without approximation.

\noindent We note that
\begin{equation}
{D^2\bmth{\xi} \over D t^2}
+ 2\bmth{\Omega_P}\times {D \bmth{\xi} \over D t} =  {\bf a}.
\label{emfs}
\end{equation}

\begin{equation}
 {\bf a} = \Delta {\bf F}
-\bmth{\Omega_P}\times \left (\bmth{\Omega_P}\times  \bmth{\xi}\right )
\end{equation}
Using equations (\ref{1sst}) and (\ref{Feq} - \ref{adiab}),
after some algebra, we can express ${\bf a}$ in the form

$${\bf a} = -{H\over \rho}\nabla \left({P' \over H}\right)
+{\nabla P\over  \rho}\bmth{\xi}\cdot \left({\nabla P\over \Gamma P}-
{\nabla \rho\over  \rho}\right) $$
\begin{equation}
\ \ \ \ \ \ \ \ \ \ \ \ \ \ \ \
 +\bmth{\xi}\cdot \nabla ({\bf v}\cdot \nabla {\bf v}
+ 2\bmth{\Omega_P}\times {\bf v}),
\label{1srhsem} \end{equation}
where $H = P^{1/\Gamma}.$

\noindent Using equations (\ref{Feq} - \ref{adiab}) we may also write
\begin{equation}
P' = -{\Gamma P \over H}\nabla (\cdot\bmth{\xi}H)
\label{presp0}\end{equation}
In what follows we find it convenient to 
decompose $\bmth{\xi}H$ into solenoidal and irrotational parts such that
\begin{equation}
\bmth{\xi}H = {\nabla \times {\bf A}\over \lambda}
+{\nabla \phi \over \lambda^3}.
\label{xidecomp}\end{equation}
The powers of the large parameter
$\lambda$ are introduced  so as to anticipate the order
of terms of different types. Thus it will  turn out that
$\bmth{\xi}H$ is mainly solenoidal with an irrotational 
correction smaller in magnitude by a factor of
order $\lambda^{-2}.$
From equation(\ref{presp0}) it then follows that
\begin{equation}
P' = -{\Gamma P \nabla^2 \phi \over H \lambda^3}
\end{equation}
and

$${\bf a} = {H\over \rho}
\nabla \left( {\Gamma P \nabla^2 \phi \over H^2 \lambda^3}\right)
+{\nabla P\over  \rho}\bmth{\xi}\cdot \left({\nabla P\over \Gamma P}-
{\nabla \rho\over \rho}\right) $$
\begin{equation}
\ \ \ \ \ \ \ \ \ \ \ \ \ \ \ \
+\bmth{\xi}\cdot  \nabla({\bf v}\cdot \nabla {\bf v}
+ 2\bmth{\Omega_P}\times {\bf v}).
\label{2srhsem} \end{equation}

\subsubsection{Effect of the Equation of State}
\noindent We comment here that the above analysis
assumed an adiabatic condition for which the stability problem
has a self-adjoint character (eg. Lynden-Bell \& Ostriker 1967).
However, because of the local nature of the formalism considered
here we may also consider other equations of state for which
the formal self-adjoint character does not hold in the fully global problem
while it is recovered in the local problem (see section \ref{govsol}).

\noindent In particular, with a later application to disk configurations
in mind we consider equations of state of the form $P =\rho c^2({\bf r}).$
Here $c$ is the sound speed which is specified to be an arbitrary function of 
${\bf r}.$ The equation of state also holds for perturbations
so that $P' = \rho' c^2.$

\noindent Then in place of equation (\ref{presp0}) we have 
\begin{equation}
P' = -c^2\nabla \cdot ( \rho \bmth{\xi}).
\label{presp1}\end{equation}
This is of the same form if we identify $H=\rho,$ and $\Gamma P = Hc^2.$
 With this identification we now get instead of equation (\ref{1srhsem}) that
\begin{equation}{\bf a} = -{Hc^2\over \rho }
\nabla \left( { P' \over H c^2 }\right)
+\bmth{\xi}\cdot  \nabla({\bf v}\cdot \nabla {\bf v}
+ 2\bmth{\Omega_P}\times {\bf v}).
\label{4srhsem} \end{equation}
This differs by the positioning of $c^2$ and the
the absence of the  second  entropy gradient term.
However, the positioning of $c^2$ is immaterial for the local analysis
presented (it may be taken through the differentiation) and thus
the formalism is the same 
as in the  adiabatic
equation of state case when the 
unperturbed 
entropy  is constant  or equivalently
$ \nabla P/(\Gamma P) = \nabla \rho /\rho.$

\subsection{Solution  Procedure in the Local Limit} \label{Local}
Although we introduced the small parameter $\lambda^{-1},$ above
we have not yet used this smallness to simplify the stability problem.
We here reduce the equation (\ref{emfs})  
together with equation (\ref{2srhsem}) to
leading order in $\lambda^{-1},$ which is here equivalent
to taking the limit $\lambda^{-1} \rightarrow 0.$

\noindent We  begin by noting that to leading order in $\lambda^{-1},$
$\bmth{\xi}H$ is solenoidal ( $\nabla\cdot (\bmth{\xi}H) = 0$)
as follows from considering  the right hand side
of (\ref{xidecomp}). 

\noindent To specify a  form for $\bmth{\xi}H$ that is  solenoidal
to leading  order in $\lambda^{-1},$
 we note that to this order only variations of $S$
need to be considered (see section \ref{loco}) so that  this is equivalent to
requiring that ${\bf k}\cdot \bmth{\xi} =0.$

\noindent In this case $\bmth{\xi}$ may be written as a linear combination
of two linearly independent 
unit vectors  ${\bf b}_1$ and  ${\bf b}_2$
that are perpendicular to ${\bf k} \propto \nabla S.$

\noindent To specify two such unit vectors we 
first use the fact that ${\bf v }\cdot \nabla S = 0$ and  adopt
\begin{equation} {\bf b}_1 = {{\bf v } \over |{\bf v }|}.
 \label{b1}\end{equation}
  We can then  adopt

\begin{equation}{\bf b}_2 = {{\bf v } \times \nabla S \over
|{\bf v } \times \nabla S|} \label{b2} \end{equation} 
which gives a unit vector orthogonal to both ${\bf v }$ and ${\bf k}.$

\noindent In adopting the above basis we make the
assumption that there are no stagnation points in the flow
at which $|{\bf v }|=0.$

\noindent
 In this case may then adopt the decomposition
\begin{equation}
\bmth{\xi} = \alpha {\bf b}_1 + \beta {\bf b}_2,
\label{fix}\end{equation}
where
$ \alpha $ and $\beta $ are of the general form given
by equation (\ref{locco}).  

\noindent The solution strategy is then to substitute
the form of the displacement (\ref{fix}) into 
equation (\ref{emfs})
together with equation (\ref{2srhsem}), to take
the scalar product of the result with respect to
${\bf b}_1$ and ${\bf b}_2$ respectively.
This results in two equations to determine
$ \alpha $ and $\beta $ on the streamline on which the perturbation
is localized.

\noindent Noting that as the dependence on $\zeta_1$ and  $\zeta_2$
can be eventually factored out and dropped and
 only convective derivatives are involved, in a Lagrangian description
  we can
ultimately regard $ \alpha $ and $\beta $
as functions only of time on a fixed streamline.
  Accordingly, these form a pair of coupled ordinary differential
equations. 

\noindent We begin by applying this procedure ro the left hand
side of equation (\ref{emfs}) which yields after straightforward
algebra:
\begin{equation}
{\ddot {\alpha}} + 2C_0{\dot {\beta}} +C_{11}\alpha + C_{12}\beta = {\bf a}
\cdot {\bf b}_1 \label{alpeq}
\end{equation}
and
\begin{equation}
{\ddot {\beta}} - 2C_0{\dot {\alpha}} +C_{21}\alpha + C_{22}\beta = {\bf a}
\cdot {\bf b}_2 . \label{beteq}
\end{equation}
Here in order to condense the notation, we
set for the first and second order convective derivatives
operating on any quantity $Q$
\begin{equation}{DQ\over Dt} \equiv {\dot Q}\end{equation}
and
\begin{equation}{D^2Q \over Dt^2} \equiv {\ddot Q}.\end{equation}
It is particularly useful to recall that for time independent quantities
such as the unit vectors introduced above
\begin{equation}{DQ\over Dt} = {\bf v}\cdot \nabla Q.\end{equation}
We further have
\begin{equation}
C_0 ={\dot {\bf b}}_2\cdot {\bf b}_1 + (\bmth{\Omega_P}\times  {\bf b}_2)
\cdot {\bf b}_1
=-{\dot {\bf b}}_1\cdot {\bf b}_2 - (\bmth{\Omega_P}\times  {\bf b}_1)
\cdot {\bf b}_2,
\label{c0}
\end{equation}
and for $i = 1,2$ and $j = 1,2$
\begin{equation}
C_{ij} ={\ddot {\bf b}}_j\cdot {\bf b}_i + (2\bmth{\Omega_P}\times
{\dot  {\bf b}}_j)
\cdot {\bf b}_i.
\label{cij}
\end{equation}

\noindent Using equation( \ref{2srhsem})  we can find
${\bf a}
\cdot {\bf b}_1,$
and
${\bf a}
\cdot {\bf b}_2 $ which occur on the right hand
sides of equations (\ref{alpeq})  and (\ref{alpeq}) respectively.

\noindent We begin by noting that  in the limit
$\lambda^{-1}  \rightarrow 0,$ $\nabla$ reduces to multiplication
by  $i\lambda \nabla S.$ Thus in this limit

$${\bf a} =
-i\left( {\Gamma P |\nabla S|^2 \nabla S \over \rho H}\right)\phi
+{\nabla P\over  \rho}\bmth{\xi}\cdot \left({\nabla P\over \Gamma P}-
{\nabla \rho\over \rho}\right) $$
\begin{equation}
\ \ \ \ \ \ \ \ \ \ 
+\bmth{\xi}\cdot  \nabla({\bf v}\cdot \nabla {\bf v}
+ 2\bmth{\Omega_P}\times {\bf v}),
\label{3srhsem} \end{equation}
with the form of $\bmth{\xi}$ being given by equation (\ref{fix}).

\noindent Using the above we find
\begin{equation}
{\bf a}
\cdot {\bf b}_1 =  D_{11}\alpha + D_{12}\beta
 \end{equation}

and
\begin{equation}
{\bf a}
\cdot {\bf b}_2 = D_{21}\alpha + D_{22}\beta,
 \end{equation} 
where for $i = 1,2$ and $i = 1,2$
$$D_{ij} = {\bf b}_i\cdot ({\bf b}_j\cdot  \nabla({\bf v}\cdot \nabla {\bf v}
+ 2\bmth{\Omega_P}\times {\bf v}))$$
\begin{equation}
\ \ \ \ \ \ \ \ \ \
+{\bf b}_i\cdot
{\nabla P\over  \rho}{\bf b}_j\cdot \left({\nabla P\over \Gamma P}-
{\nabla \rho\over \rho}\right). \end{equation}

\noindent Thus the governing equations for $\alpha$
and $\beta$ may be cast in the final form
\begin{equation}
{\ddot {\alpha}} + 2C_0{\dot {\beta}} =E_{11}\alpha + E_{12}\beta 
\label{alpeqf}
\end{equation}
and
\begin{equation}
{\ddot {\beta}} - 2C_0{\dot {\alpha}} =E_{21}\alpha + E_{22}\beta,
 \label{beteqf}
\end{equation}
where $E_{i,j} = D_{i,j}-C_{i,j}.$

\noindent To complete the solution the scalar $\phi$
 may be determined by taking
the scalar product of equation(\ref{emfs}) with ${\bf k}.$
However, this is not needed for our purposes so  we do not consider it further
here.

\noindent We comment that the following identities, valid for
any unit vectors, are useful for evaluating the coefficients in the governing equations
for $\alpha$ and $\beta$ that involve second time derivatives:
\begin{equation}
{\ddot{\bf b}}_i\cdot {\bf b}_i= - |\dot {\bf b}_i|^2 \label{iden1}\end{equation}
\noindent and
\begin{equation}
{\ddot{\bf b}}_i\cdot {\bf b}_j=-\dot{\bf b}_i\cdot \dot{\bf b}_j
+ {D (\dot{\bf b}_i\cdot {\bf b}_j)\over Dt} \label{iden2}\end{equation}

\noindent The governing equations (\ref{alpeqf}- \ref{beteqf})
consist of a pair of second order differential equations
for the amplitudes $(\alpha, \beta)$ as a function of time.
As the above is a Lagrangian description the coefficients appearing in these
equations generically are periodic functions of time.
This is because these coefficients describe fluid effects
as seen by a fluid element as it circulates  periodically on an unperturbed
streamline.  Only when these conditions are invariant 
 such as on circular streamlines in a  system
  with cylindrical symmetry are the coefficients constant.
In the generic case the periodicity of the coefficients
leads to a system similar to the well known cases of
 Mathieu's or Hill's equation
(eg. Whittaker \& Watson 1996). Accordingly we expect to encounter
parametric instability when a normal mode frequency is commensurable
with the orbital frequency associated with  circulating around a streamline.
The lowest order relation is when the normal mode frequency is
one half of the  orbital frequency. This is easy to demonstrate when the
system is almost cylindrically symmetric and the
streamlines depart from circles by a small amount (see below
and appendix 2).
Furthermore commensurability conditions can always be satisfied
when the underlying modes are gravity/inertial modes because,
as is well known, their spectrum is dense (eg. Papaloizou \& Pringle 1982;
Lin Papaloizou \& Kley 1993).
This density comes about because in the local limit
the oscillation frequencies depend only on wavenumber ratios
so that wavenumber  components may be allowed to tend to infinity
at fixed ratios   while 
not affecting the oscillation frequency.
 One can use that  property
to relate the spectrum of the local problem to that of the 
global problem ( see appendix 3).  We illustrate the above remarks
by considering a system with cylindrical symmetry and a system
for which the streamlines are Keplerian ellipses of arbitrary 
eccentricity. Of course in the latter case  when the eccentricity
is small the streamlines deviate slightly from circles.

\section {The Case of Circular Streamlines in the Steady State} \label{sime}

\noindent In this case, $\bmth{\Omega_P} = 0, $
and  the invariants on a streamline are
taken to be $\zeta_1 = r,$ and $\zeta_2 = z.$
Because the orbits are circles we have $r = a,$ where $a$
is the semi-major axis.

\noindent For the phase function we adopt  $S =K_r r +K_z z$
with $K_r$ and $K_z$ being constants.

\noindent This gives the wavenumber
${\bf k} = \lambda \nabla S =  (\lambda K_r, 0, \lambda K_z).$

\noindent For this steady state model ${\bf v}= (0, v_{\varphi}(r), 0),$
equation (\ref{b1}) gives ${\bf b}_1 ={\hat {\bmth {\varphi}}}$ being the unit vector
in the azimuthal direction. Then from equation
(\ref{b2})  we obtain ${\bf b}_2 = (K_z {\hat {\bf r}} - K_r {\hat {\bf k}})/K,$
where ${\hat {\bf r}}$ and ${\hat {\bf k}}$ are unit vectors
in the radial and vertical directions respectively
and $K =\sqrt{K_r^2 + K_z^2}.$

\noindent    Having obtained these   unit vectors the forms of
equations (\ref{alpeqf}) and (\ref{beteqf}) can be found
by direct calculation.

\noindent    Using $D{\hat {\bf r}}/Dt = {\bf v}\cdot\nabla{\hat {\bf r}} =  v_{\varphi}
{\hat {\bmth{\varphi}}}/r,$ and $D{\hat{\bmth{\varphi}}}/Dt  = {\bf v}\cdot\nabla{\hat 
{\bmth{ \varphi}}}
= -v_{\varphi}{\hat {\bf r}}/r,$ we find
setting $\Omega = v_{\varphi}/r $ which
is constant on streamlines,  that
\begin{equation}
{D {\bf b}_1 \over Dt} = -\Omega{\hat {\bf r}}
\end{equation}
and
\begin{equation}
{D {\bf b}_2 \over Dt} = 
{K_z \Omega {\hat {\bmth{\varphi}}}\over K}.
\end{equation}
From these it is easy to obtain
$ C_0 = {\bf b}_1\cdot {\dot{\bf b}}_2 = K_z\Omega/K,$
${\ddot{\bf b}}_1\cdot {\bf b}_2 ={\ddot {\bf b}}_2\cdot {\bf b}_1 = 0,$
${\ddot{\bf b}}_1\cdot {\bf b}_1 = -\Omega^2,$ and 
${\ddot{\bf b}}_2\cdot {\bf b}_2 = -K_z^2\Omega^2/K^2.$

\noindent Further we find that for $\bmth{\xi}$ given by equation (\ref{fix}) that
\begin{equation}
 {\bf b}_1\cdot (\bmth{\xi} \cdot  \nabla({\bf v}\cdot \nabla {\bf v}))
= -\Omega^2 \alpha ,\label{cf0}
\end{equation}
and
\begin{equation}
 {\bf b}_2\cdot (\bmth{\xi} \cdot  \nabla({\bf v}\cdot \nabla {\bf v}))
= -{\Omega^2 \beta K_z^2\over K^2} -
{2\Omega r\Omega' \beta K_z^2\over K^2}.
\label{cf1}
\end{equation}
In addition for $i=1,2 ,j =1,2 ,$
$$ {\bf b}_i\cdot
{\nabla P\over  \rho}{\bf b}_j\cdot \left({\nabla P\over \Gamma P}-
{\nabla \rho\over \rho}\right)=  {\rm \hspace{4cm}   }$$
\begin{equation}{{\hat {\bmth{\varphi}}}\cdot ({\bf k}\times \nabla P)\over |{\bf k}|\rho}
{{\hat {\bmth{\varphi}}}\cdot\left({\bf k}\times \left({\nabla P\over \Gamma P}-
{\nabla \rho\over \rho}\right)\right)\over |{\bf k}|} \delta_{i2} \delta_{j2},
\label{cf2}\end{equation}
with $ \delta_{ij}$ denoting the Kronnecker delta.

\noindent Using (\ref{cf0}-\ref{cf2}) and  above,
equations (\ref{alpeqf}) and (\ref{beteqf})
are  reduced to the simple forms 

\begin{equation}
{\ddot {\alpha}} + {2K_z\Omega\over K}{\dot {\beta}} =0
\label{alpeqsim}
\end{equation}
 \begin{equation}
{\ddot {\beta}} - {2K_z\Omega\over K}{\dot {\alpha}}  =-\left(
{2\Omega r\Omega' K_z^2\over K^2} +  \nu^2\right) \beta,\end{equation}
where
\begin{equation}\nu^2 K^2  =
-{\hat {\bmth{\varphi}}}\cdot\left({\bf k}\times \left({\nabla P\over \rho}\right)\right)
{\hat {\bmth{\varphi}}}\cdot\left({\bf k}\times \left({\nabla P\over \Gamma P}-
{\nabla \rho\over \rho}\right)\right)
 \label{beteqsim}
\end{equation}

\noindent In this case because there is no variation of state
variables around streamlines the governing equations
have constant coefficients and we may look for normal mode
solutions for which $\alpha $ and $\beta $  are $\propto \exp(i\sigma t)$
where $\sigma$ is the normal mode frequency.

\noindent Setting $\alpha  = \alpha _0  \exp(i\sigma t)
, \beta = \beta_0 \exp(i\sigma t)$ equations (\ref{alpeqsim}-\ref{beteqsim})
give a non trivial solution such that
\begin{equation}
\alpha = {2iK_z\Omega\over K\sigma }\beta  \end{equation}
and 
\begin{equation}
\sigma^2 ={ 2\Omega (r\Omega' +2\Omega) K_z^2 \over K^2}  + \nu^2 
. \label{kepinert}\end{equation}

This gives the well known local dispersion for inertial/gravity modes
in a differentially rotating system (see eg. Tassoul 1978 and references
therein for a discussion). There are two contributions. The first 
is from the rotation and depends on the 
gradient of angular momentum which must be positive for stability
(Rayleigh's criterion) and the second is from the entropy gradient
and leads to gravity waves when stable ($\nu^2 > 0$)  or convection when convectively
unstable (Schwarzchild's criterion ). 
 We shall always assume that the system is  stable to convection
and satisfies Rayleigh's criterion. Accordingly we consider 
systems that are  dynamically stable in this sense.

\noindent However, as we shall see below parametric instability can occur
when state variables vary around streamlines making a perturbed
fluid element subject to periodic restoring forces even though
the system is  stable according to the Rayleigh and Schwarzchild
criteria.

\noindent An important feature in this context is that the oscillation
spectrum is dense. This is because the local dispersion relation depends
only on the wavenumber ratio $K_r/K_z.$  As in a local analysis
both $\lambda K_r$ and $\lambda K_z$ $\rightarrow \infty$ these may be adjusted 
to make the ratio take on any finite value and thus produce
a continuous or everywhere dense spectrum
(see eg. Papaloizou \& Pringle  1982; Lin et al 1993 for
a discussion of this point in a global context).

\noindent This property of the spectrum means that commensurability conditions
such as $\sigma = \Omega/2,$ with $\Omega$ being the angular frequency
associated with one orbit around a streamline, required for parametric
instability can always be  adequately satisfied. 

\noindent For example 
(\ref{kepinert}) gives for inertial modes $(\nu^2 =0)$
and Keplerian rotation, $\sigma^2 = \Omega^2 K_z^2/K^2.$ 
Thus when $K_z/K_r = 1/ \sqrt{3},$ $\sigma = \pm \Omega /2.$

\noindent Suppose, the radial and vertical wavenumber
components have to be integral multiples
of some smallest values $(K_{r0} , K_{z0})$ respectively. Then
 because we can let both wavenumber components become
arbitrarily large, and because of the ability
to approximate irrationals arbitrarily closely with rationals,
we can  make $|K_z/K_r - \sqrt{3}|$ become as small
as we like.

\section {A steady state model with streamlines being Keplerian 
ellipses} \label{notsim}
We now consider a more interesting non axisymmetric model
for which the streamlines correspond to Keplerian ellipses with the same eccentricity
circulating about a central point mass. This is described in section
\ref{Stmodel}. In this particular model the streamlines form a pattern
that is stationary in the inertial frame.
  However,  in more general cases the pattern may precess
slowly and so appear stationary only in a slowly rotating frame.
But this effect  should make little difference.
We comment also that because we are considering local stability,
where perturbations are localized along streamlines, these streamlines
are only required to fill a small local volume of the fluid
independently of  exterior conditions.

\noindent In this   model fluid state variables
vary as a fluid element circulates around a streamline. Thus
periodic forcing of a perturbed fluid element occurs
and parametric instability becomes possible.

\noindent The velocity field for the  streamlines  was specified
through equations (\ref{radvel}) - (\ref{phivel})
in section \ref{Stmodel}. To proceed we need
to specify two invariants  for the streamlines.
Because for the cylindrical disk model adopted,
the motion is in horizontal planes so one of these may be taken to be
$z.$ The other may be taken to be $a(1-e^2) = r(1 + \cos \varphi))$
which is proportional to the square of the angular momentum.

\begin{figure}
\begin{center}
\epsfig{file=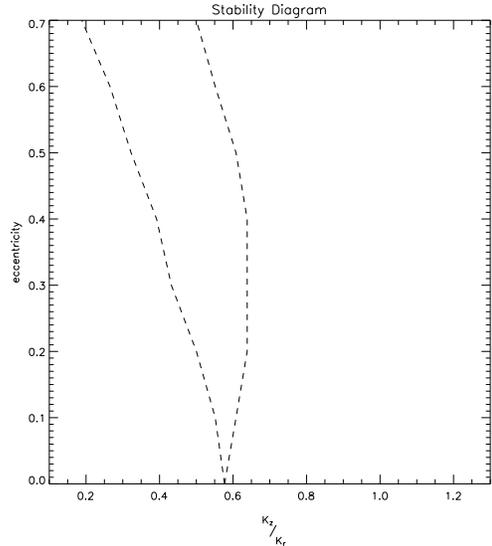, height=7cm, width=7cm, angle=360}
\caption
{The stability diagram for equations
(\ref{alpeqf}) and (\ref{beteqf}) applied to the model
with streamlines  which are  Keplerian ellipses in
the $( {K_z\over K_r}, e)$ plane.
Instability occurs in the region of the  plane  between the two curves plotted.   
\label{fig1}}
\end{center}
\end{figure}

\begin{figure}
\begin{center}
\epsfig{file=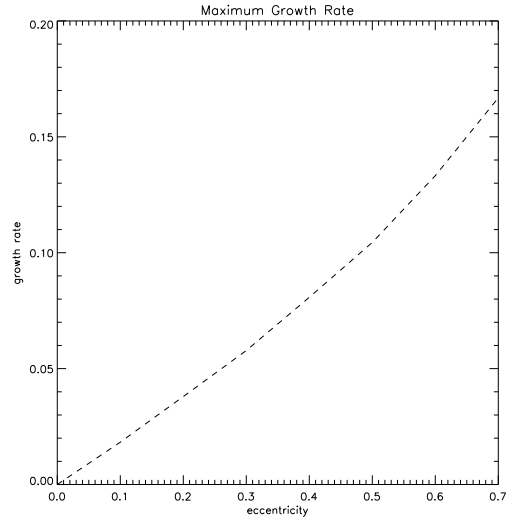, height=7cm, width=7cm, angle=360}
\caption{ Maximum growth rate in units of $\sqrt{GM_*/a^3}$
plotted as a function of eccentricity for solutions of equations
(\ref{alpeqf}) and (\ref{beteqf})
applied to the model
with streamlines  which are  Keplerian ellipses.  
\label{fig2}}
\end{center}
\end{figure}

\noindent   Using these we adopt
for the phase function

\noindent $S =K_r r(1 +e \cos\varphi) + K_z z$
with $K_r$ and $K_z$ again  being constants.

\noindent The wavenumber is  then given by

\noindent ${\bf k} = \lambda \nabla S = (\lambda K_r(1 +e \cos\varphi),
 -\lambda K_r e\sin\varphi, \lambda K_z).$

\noindent In this case
\begin{equation} {\bf v}= \left(
\sqrt{{GM_*\over r(1+e\cos\varphi)}} e\sin\varphi,
\sqrt{{GM_*(1+e\cos\varphi)\over r}},
 0\right),\end{equation}
so that 
equation (\ref{b1}) gives 
\begin{equation}
{\bf b}_1 ={{\bf v}\over |{\bf v}|} =  
{ e\sin\varphi {\hat{\bf r}}+(1 +  e\cos\varphi){\hat{\bmth{\varphi}}}
 \over \sqrt{(1+ e^2 + 2e\cos\varphi)}}
\end{equation}

\noindent From equation
(\ref{b2}) we obtain 
 $$ {\bf b}_2 = 
{K_z(1 +  e\cos\varphi){\hat{\bf r}}-K_z e\sin\varphi{\hat{\bmth{\varphi}}}
\over N \sqrt{(1+ e^2 + 2e\cos\varphi)}}$$
\begin{equation}  \ \ \ \ \ \ \ \ \ \ \ \
 -{K_r\sqrt{(1+ e^2 + 2e\cos\varphi)}{\hat{\bf k}}
\over N},\end{equation}
where  ${\hat {\bf k}}$ is the  unit vector
in the vertical direction 
and $N =\sqrt{K_r^2(1+ e^2 + 2e\cos\varphi) + K_z^2}$
is the normalizing factor required to ensure that ${\bf b}_2$ is a unit vector.

\noindent   Given these  unit vectors the forms of
equations (\ref{alpeqf}) and (\ref{beteqf}) can again  be found
by direct calculation. As the details are somewhat tedious they are 
relegated to  appendix 1.

\noindent  The
 coefficients $( E_{i,j}, i=1,2, j=1,2, C_0)$
  contained in the linear governing equations
 (\ref{alpeqf}) and (\ref{beteqf}) for $\alpha$ and $\beta$ are given by 
$$E_{11} = - {GM_*\over r^3}{{(1+2e\cos\varphi +e^2(1-3\sin^2\varphi))}
\over 1+ e^2 + 2e\cos\varphi}+\hspace{6cm}$$
\begin{equation}
{GM_*\over r^3}{(1+e\cos\varphi)^{3}\over (1+ e^2 + 2e\cos\varphi)^{2}},
\end{equation}
\begin{equation}
E_{12} ={GM_*\over r^3}{3K_z e\sin\varphi (1+e\cos\varphi)
\over N(1+ e^2 + 2e\cos\varphi)}
+{\dot{\bf b}}_1\cdot \dot{\bf b}_2
- {D C_0\over Dt}\nonumber ,\end{equation}
\begin{equation}
E_{21} ={GM_*\over r^3}{3K_z e\sin\varphi (1+e\cos\varphi)
\over N(1+ e^2 + 2e\cos\varphi)}
+{\dot{\bf b}}_1\cdot \dot{\bf b}_2
+ {D C_0\over Dt}\nonumber  \end{equation}
and
\begin{equation}
E_{22} =  {GM_*\over r^3}{{K_z^2(2+4e\cos\varphi +e^2(2-3\sin^2\varphi))}
\over N^2 ( 1+ e^2 + 2e\cos\varphi)} - {\bf b}_2 \ddot{\bf b}_2.
\end{equation}

\noindent   In addition
\begin{equation} C_0 = {\bf b}_1\cdot {\dot{\bf b}}_2 =
\sqrt{GM_*\over r^3}{K_z (1+e\cos\varphi)^{3/2}
\over N (1+ e^2 + 2e\cos\varphi)} \label{start}
\end{equation}

\noindent  and
$${D C_0 \over Dt} =
-{GM_*\over r^3}{K_z(1+e\cos\varphi)e\sin\varphi\over N(1+ e^2 + 2e\cos\varphi)}\times
\hspace{6cm} $$
\begin{equation} \left({(1 + 3e^2 +4 e\cos\varphi)\over  (1+ e^2 + 2e\cos\varphi)}
-{K_r^2 (1+e\cos\varphi) \over N^2} \right)\end{equation}

\noindent with 

\begin{equation} {D N \over Dt}
= -\sqrt{GM_*\over r^3}
{ K_r^2 e\sin\varphi\sqrt{1 +  e\cos\varphi} 
\over \sqrt{ K_z^2 + K_r^2 (1+ e^2 + 2e\cos\varphi)}}.
\end{equation}

\noindent   we also have
$${\ddot{\bf b}}_2\cdot {\bf b}_2 =- 
{GM_*(1 +  e\cos\varphi)\over r^3} \times \hspace{6cm}$$
$$ \left({{(1 +  e\cos\varphi)^2K_z^2+K_r^2e^2\sin^2\varphi
(1+e^2+2e\cos\varphi)}\over N^2 (1+e^2+2e\cos\varphi)^2} \right. \hspace{6cm}$$
\begin{equation} \left. -{{K_r^4e^2\sin^2\varphi}\over 
N^2 (K_z^2 + K_r^2 (1+ e^2 + 2e\cos\varphi))}\right) .\end{equation}

\noindent together with

\begin{equation}{\dot{\bf b}}_2\cdot \dot{\bf b}_1
=\sqrt{GM_*\over r^3}{DN\over Dt}
{K_z (1+e\cos\varphi)^{3/2}
\over N^2 (1+ e^2 + 2e\cos\varphi)}.
\end{equation}
which enables all the coefficients to be calculated.

\section{Solution of the governing equations}\label{govsol}
\noindent  We  first  note that the  governing
equations~(\ref{alpeqf}-\ref{beteqf}) can be written in a  compact vector form
with similar structure to the Lynden-Bell \& Ostriker (1967) formulation
of the general stability problem (see equations (\ref{1sem}) and (\ref{LBO})):
\begin{equation}
{\bf \ddot y} -2C_0{\bf e_3}\times {\bf \dot y}
= {\bf E}\cdot {\bf  y} .\label{LBO1}
\end{equation}
Here the two dimensional vector $ {\bf  y}^T = (\alpha , \beta)$ and
the vector ${\bf e_3}\times {\bf \dot y} = (-\dot \beta , \dot \alpha)^T.$
The  two dimensional matrix ${\bf E}$ has elements
$E_{11}, E_{12}, E_{21}, E_{22}$ as defined  in section \ref{Local}
above.

\noindent When a system has stream lines  perturbed slightly
from circles and the unperturbed circular streamline case has 
an underlying spectrum of oscillation modes with a dense spectrum,
it is possible to show analytically that the system has a generic parametric
instability. This is done in appendix 2.

\noindent However, we have found that the most direct method of assessing the 
stability of a general system, with streamlines
differing significantly from circles, governed by equation (\ref{LBO1}) is to
integrate the system as an initial value problem starting from general initial 
conditions  and to examine whether the solutions manifest exponential growth (se below).
In this way, which is effectively exploiting the Lagrangian 
nature of the system and  solving for  conditions
in the neighbourhood of a fluid element as it circulates around
streamline,  problems with boundary conditions are avoided.

\noindent However, before  presenting an example we briefly comment that
the system (\ref{LBO1}) can be dealt with in an equivalent  Eulerian formulation in which
an eigenvalue problem of the same type as that in the Lynden-Bell \& Ostriker formalism
is recovered. To obtain the Eulerian formulation the
convective derivatives in (\ref{LBO1}) are replaced 
as in (\ref{svconv}). Thus

\begin{equation}{D  \bmth{y} \over D t} =
{\partial \bmth{y}\over \partial t}+ {\bf v}\cdot \nabla  \bmth{y}.
\label{svconvn}.\end{equation}
As our system has spatial variation only around streamlines, we can write
${\bf v}\cdot \nabla  \bmth{y} =|{\bf v}| d\bmth{y}/ds,$ with $s$
measuring the arc length along a streamline. The problem is thus a one dimensional one
with periodic boundary conditions in $s.$
One now adopts a time dependence through a factor $\exp(\i\sigma t)$
and obtains an eigenvalue problem for  $\sigma$  of the form (\ref{LBO})
namely
\begin{equation}
-\sigma^2\bmth{y} +\sigma {\cal B}(\bmth{y})
+ {\cal C}(\bmth{y}) = 0, \label{LBO2}
\end{equation}
where each of the operators ${\cal B}, {\cal C}$ is self adjoint
with the inner product for general vectors ${\bf y}_1, {\bf y}_2$
defined as
\begin{equation}
({\bf y}_2, {\bf y}_1) = \int \rho {\bf y}_2^*\cdot {\bf y}_1 {\cal A}ds,
\end{equation}
where ${\cal A}$ is the stream tube area function
(see section \ref{stream}) and  the integral is taken around the  streamline of interest.

\noindent This formulation allows us to note some of the results of Lynden-Bell
and Ostriker (1967) such as that eigenvalues occur in quartets with complex conjugates
and with opposite signs.

\subsection{Numerical solution for Kepler Ellipses}
 We have solved the governing 
equations (see equation (\ref{LBO1}) in section \ref{govsol} numerically as an initial value 
problem. To do this we specify the ratio $K_z/K_r$ and integrate the system forward in time
from arbitrary initial conditions.
 Instability then shows up as exponential growth, the rate may be easily estimated in practice.
This procedure is very much simpler than 
attempting to solve a matrix eigenvalue problem. 

\noindent In the limit of small eccentricity the problem can be solved
analytically using a standard approach to parametric instability
problems. This is done in appendix 2.
In this case the maximum  instability occurs when $K_z/K_r =1/\sqrt{3}.$
At this wavenumber ratio, the maximum growth rate is
\begin{equation} \lambda = 3e\Omega /16. \label{conds}\end{equation}

\noindent The calculation is a Lagrangian one so viewed from the point of view of a fluid element
circulating around a streamline, the instability occurs through the
interaction of two inertial modes with frequencies $\pm \Omega /2.$ 
In an Eulerian sense, if the pattern is stationary in the inertial frame,
the frequencies as viewed there are $-m_1\Omega +\Omega /2$ and $-m_2\Omega -\Omega /2,$
where $(m_1,m_2)$ are the associated azimuthal mode numbers. These frequencies
must be equal in the stationary  frame because there is only one eigenfrequency
$\sigma$ as viewed from there as considered above.
This means that $m_1 - m_2 = 1.$ Otherwise the $m_i$ are not constrained. 

Numerical solution of the governing equations
gives the stability diagram 
plotted in figure \ref{fig1}. This shows the region of
the $(K_z/K_r , e )$ plane with instability for $0 < e < 0.7.$
When $e \rightarrow 0,$ the numerical and analytic results agree.
The region of instability collapses onto $K_z/K_r =1/\sqrt{3}$
and the maximum growth rate  agrees with equation (\ref{conds}).
The band of instability increases with $e.$

\noindent The Maximum growth rate in units of $\sqrt{GM_*/a^3}$
is plotted as a function of eccentricity 
in figure \ref{fig2}. This indicates that the maximum growth rate
for a given $e,$ is such that, to a good approximation, as in the small $e$ limit, 
$\lambda \propto e$
for $e < 0.7 .$

\section{Discussion}\label{Discu}
In this  paper we have formulated a general local
stability analysis 
applicable to astrophysical flows that appear  steady 
in an appropriate uniformly rotating coordinate system.
Such a  flow might correspond to a stellar mode of oscillation 
that is a travelling wave in the azimuthal direction or an
accretion disk with a  free eccentricity. In the simple
example we considered  
the disk streamlines were  Keplerian ellipses.
 
\noindent The general stability analysis considered perturbations
localized in the neighbourhood of unperturbed fluid streamlines.
The spatial localization was assumed to tend to zero as the wavenumber
tended to infinity but more slowly than the inverse magnitude of 
the wavenumber itself
allowing the adoption of wave packets. It is possible 
to consider the limit of infinite wave number taken at finite
oscillation eigenfrequency and obtain spectral properties
from the local theory that apply to the global problem.
The physical reason for this is that the degrees of freedom
under consideration correspond to inertial or gravity modes
whose frequency depends only on the ratio of wavenumber components.
The group velocity thus approaches zero with localization in
a way that ensures that localization may be maintained for arbitrarily long
times. This justifies the local procedure.

\noindent We found that the local  stability of  fluid
on a particular streamline was governed by a pair of second order
differential equations, which in a Lagrangian description,
evolved the  perturbation associated 
with a particular fluid element as a function of time. This reduces the
stability problem to an initial value problem rather than an 
equivalent eigenvalue problem obtained in an Eulerian representation.

\noindent When conditions do not vary  around streamlines
one obtains standard oscillation modes. However, in the more general case
when conditions vary around streamlines the ordinary differential
equations are linear with periodic coefficients and 
the system  was shown to 
 become  generically parametrically unstable. The dense spectra
of inertial or gravity  modes ensures that resonance conditions may always
be satisfied.

\noindent We applied our considerations to study the stability
of fluid for which the streamlines were  in Keplerian ellipses
with eccentricity up to 0.7.
We showed that
for small $e,$
the maximum  instability occurs when $K_z/K_r =1/\sqrt{3}$
and involves, in the Eulerian description, 
two modes with azimuthal mode number
differing by unity in magnitude.
Then the growth rate is
\begin{equation} \lambda = 3e\Omega /16. \end{equation}
For large eccentricities, to a good approximation,
 this gives the maximum growth rate as a 
function of eccentricity. However, the band of unstable wavenumber
ratios shifts (see figure \ref{fig1}).

\noindent Finally the considerations 
in this paper are based on linear theory and
so they cannot determine the outcome of the instability.
Results based on three dimensional
numerical simulations
for accretion disks with elliptical streamlines will
be presented in a following publication.

 
\begin{appendix}\label{A.1}
\title{{\bf APPENDIX 1}}
\section*{Keplerian streamline model}
We here consider the derivation of the coefficients
in the linear governing equations for $\alpha$ and
$\beta$ for the model with streamlines corresponding to
Keplerian ellipses described in section (\ref{notsim}).
The velocity field  is given by
equations (\ref{radvel}) - (\ref{phivel})
and the two invariants  on streamlines
are taken to be
$z$  and $a(1-e^2) = r(1 + \cos \varphi))$
being proportional to the square of the angular momentum.

\noindent We here note that the well known orbit equation
given above may be used to eliminate $r$ in the expressions
given below
so that the coefficients only depend on $\varphi$
and quantities constant on a streamline.
Given the dependence
of $\varphi$ on time, 
they may also be expressed as functions of time and
quantities constant on a streamline.

\noindent  The phase function is

\noindent $S =K_r r(1 +e \cos\varphi) + K_z z$
with $K_r$ and $K_z$  being constants.

\noindent The wavenumber is given by

\noindent ${\bf k} = \lambda \nabla S =  (\lambda K_r(1 +e \cos\varphi),
 -\lambda K_r e\sin\varphi, \lambda K_z).$

\noindent In this case
\begin{equation} {\bf v}= \left(
\sqrt{{GM_*\over r(1+e\cos\varphi)}} e\sin\varphi,
\sqrt{{GM_*(1+e\cos\varphi)\over r}},
 0\right),\end{equation}
so that 
equation (\ref{b1}) gives 
\begin{equation}
{\bf b}_1 ={{\bf v}\over |{\bf v}|} =  
{ e\sin\varphi {\hat{\bf r}}  +(1 +  e\cos\varphi){\hat{\bmth{\varphi}}}
\over \sqrt{(1+ e^2 + 2e\cos\varphi)}} \label{b1a}
\end{equation}

\noindent From equation
(\ref{b2}) we obtain 
 $$ {\bf b}_2 = 
{ K_z(1 +  e\cos\varphi){\hat{\bf r}} -K_z e\sin\varphi {\hat{\bmth{\varphi}}} 
\over N \sqrt{(1+ e^2 + 2e\cos\varphi)}} \hspace{1cm}
$$
\begin{equation}  \ \ \ \ \ \ \ \ \ \ \ \
 -{K_r\sqrt{(1+ e^2 + 2e\cos\varphi)}{\hat{\bf k}}
\over N},\label{b2a}\end{equation}
where  ${\hat {\bf k}}$ is the  unit vector
in the vertical direction 
and $N =\sqrt{K_r^2(1+ e^2 + 2e\cos\varphi) + K_z^2}$
is the normalizing factor required to ensure that ${\bf b}_2$ is a unit vector.

\noindent   Given these  unit vectors the forms of
equations (\ref{alpeqf}) and (\ref{beteqf}) can again  be found
by direct calculation.

\noindent  For this purpose it is useful to note for this model
that  the operator
$$ {D \over Dt} \equiv  {\bf v}\cdot \nabla \hspace{8cm}
$$
\begin{equation} = \sqrt{GM_*\over r^3}
\left ( \sqrt{(1+e\cos\varphi)}{\partial \over \partial \varphi}
+{re\sin\varphi\over \sqrt{(1+e\cos\varphi)}}
 {\partial  \over  \partial r} \right)  \end{equation}

\noindent and  to recall that

\noindent $\partial {\hat {\bf r}} / \partial \varphi = 
{\hat {\bmth{\varphi}}} $ and 
 $\partial {\hat {\bf  {\hat{\bmth{\varphi}}}}} / \partial \varphi =
- {\hat {\bf r}}, $
while the unit vectors have zero derivative  with respect to $ r.$

\noindent  Applying this operator to equation (\ref{b1}) we obtain
$${D {\bf b}_1 \over Dt} = \sqrt{GM_*\over r^3}
{(1+e\cos\varphi)^{3/2}\over (1+ e^2 + 2e\cos\varphi)^{3/2}} \times \hspace{6cm}$$
\begin{equation} \left( -(1+e\cos\varphi){\hat {\bf r}}
+e\sin\varphi {\hat{\bmth{\varphi}}}\right) \label{db1dt}
\end{equation}
and applying it to equation (\ref{b2}) we get
$${D(N  {\bf b}_2 )\over Dt} =
\sqrt{GM_*\over r^3}\left(
{ K_z(1 +  e\cos\varphi)^{3/2} e\sin\varphi 
\over {(1+ e^2 + 2e\cos\varphi)^{3/2}}}{\hat{\bf r}} \right. \hspace{6cm} $$

$$+{K_z(1 +  e\cos\varphi)^{5/2} 
\over {(1+ e^2 + 2e\cos\varphi)^{3/2}}}{\hat{\bmth{\varphi}}}\hspace{6cm}$$
\begin{equation} 
\left. +{ K_r(1 +  e\cos\varphi)^{1/2} e\sin\varphi
\over \sqrt{(1+ e^2 + 2e\cos\varphi)}}{\hat{\bf k}}\right),
\end{equation}

\noindent with 

\begin{equation} {D N \over Dt}
= -\sqrt{GM_*\over r^3}
{ K_r^2 e\sin\varphi\sqrt{1 +  e\cos\varphi} 
\over \sqrt{ K_z^2 + K_r^2 (1+ e^2 + 2e\cos\varphi)}}\label{dndt}.
\end{equation}

\noindent From the  above relations we obtain
\begin{equation} C_0 = {\bf b}_1\cdot {\dot{\bf b}}_2 =
\sqrt{GM_*\over r^3}{K_z (1+e\cos\varphi)^{3/2}
\over N (1+ e^2 + 2e\cos\varphi)} \label{Astart}
\end{equation} 

\begin{equation}{\ddot{\bf b}}_1\cdot {\bf b}_1 = -|\dot {\bf b}_1|^2 =
-{GM_*\over r^3}{(1+e\cos\varphi)^{3}\over (1+ e^2 + 2e\cos\varphi)^{2}}
,\end{equation}
and
$${\ddot{\bf b}}_2\cdot {\bf b}_2 = - |\dot {\bf b}_2|^2  = - 
{GM_*(1 +  e\cos\varphi)\over r^3} \times \hspace{6cm}$$

$$ \left({{(1 +  e\cos\varphi)^2K_z^2+K_r^2e^2\sin^2\varphi
(1+e^2+2e\cos\varphi)}\over N^2 (1+e^2+2e\cos\varphi)^2} \right. \hspace{6cm}$$
\begin{equation} \left. -{{K_r^4e^2\sin^2\varphi}\over 
N^2 (K_z^2 + K_r^2 (1+ e^2 + 2e\cos\varphi))}\right) .\end{equation}

\noindent We now use the general results for any pair of unit vectors
obtained from (\ref{iden1}-\ref{iden2})  that
\begin{equation}{\ddot{\bf b}}_2\cdot {\bf b}_1=-{\dot{\bf b}}_1\cdot \dot{\bf b}_2
+ {D C_0\over Dt}\nonumber \end{equation}
and
\begin{equation}{\ddot{\bf b}}_1\cdot {\bf b}_2=-{\dot{\bf b}}_2\cdot \dot{\bf b}_1
- {D C_0\over Dt}\nonumber \end{equation}
to obtain ${\ddot{\bf b}}_2\cdot {\bf b}_1$ and ${\ddot{\bf b}}_1\cdot {\bf b}_2$
which occur in the equations governing $\alpha$ and $\beta.$

\noindent We first 
apply the convective derivative operator to (\ref{Astart}) 
to obtain
$${D C_0 \over Dt} =
-{GM_*\over r^3}{K_z(1+e\cos\varphi)e\sin\varphi\over N(1+ e^2 + 2e\cos\varphi)}\times
\hspace{6cm} $$
\begin{equation} \left({(1 + 3e^2 +4 e\cos\varphi)\over  (1+ e^2 + 2e\cos\varphi)}
-{K_r^2 (1+e\cos\varphi) \over N^2} \right)\end{equation}
and use (\ref{db1dt} - \ref{dndt}) to obtain
\begin{equation}{\dot{\bf b}}_2\cdot \dot{\bf b}_1
=\sqrt{GM_*\over r^3}{DN\over Dt}
{K_z (1+e\cos\varphi)^{3/2}
\over N^2 (1+ e^2 + 2e\cos\varphi)}
\end{equation}
which enables these quantities to be found directly.

\noindent For this model, because the velocity field
is Keplerian, we have the very useful result that
\begin{equation}
{\bf v}\cdot \nabla {\bf v} =
-{GM_*\over r^2} {\hat{\bf r}}.\end{equation}

\noindent   Using this together with $\bmth{\xi}$
 given by equation (\ref{fix})
we obtain 
$$
 {\bf b}_1\cdot (\bmth{\xi} \cdot  \nabla({\bf v}\cdot \nabla {\bf v}))
 =  \beta {GM_*\over r^3}{3K_z e\sin\varphi (1+e\cos\varphi)
\over N(1+ e^2 + 2e\cos\varphi)}$$
\begin{equation}
-\alpha {GM_*\over r^3}{{1+2e\cos\varphi +e^2(1-3\sin^2\varphi)}
\over 1+ e^2 + 2e\cos\varphi} ,\label{cfw0}
\end{equation}
and
 $${\bf b}_2\cdot (\bmth{\xi} \cdot  \nabla({\bf v}\cdot \nabla {\bf v}))
= \alpha {GM_*\over r^3}{3K_z e\sin\varphi (1+e\cos\varphi)
\over N(1+ e^2 + 2e\cos\varphi)}$$
\begin{equation}+\beta {GM_*\over r^3}{{K_z^2(2+4e\cos\varphi +e^2(2-3\sin^2\varphi))}
\over N^2 ( 1+ e^2 + 2e\cos\varphi)}
\label{cfw1}
\end{equation}
In addition for this particular
model with uniform pressure, we automatically have  for $i=1,2 ,j =1,2 ,$
 \begin{equation} {\bf b}_i\cdot
{\nabla P\over  \rho}{\bf b}_j\cdot \left({\nabla P\over \Gamma P}-
{\nabla \rho\over \rho}\right)= 0
\label{cf5}.\end{equation}

\noindent We have now assembled all the quantities needed to
calculate the coefficients $(C_0, E_{i,j}, i=1,2, j=1,2)$
 contained in the linear governing equations 
(\ref{alpeqf}) and (\ref{beteqf}) for $\alpha$ and $\beta.$
These can be  constructed using equations (\ref{cij}-\ref{iden2})
of  section \ref{Linstab}.
We obtain:
$$E_{11} = - {GM_*\over r^3}{{(1+2e\cos\varphi +e^2(1-3\sin^2\varphi))}
\over 1+ e^2 + 2e\cos\varphi}+\hspace{6cm}$$
\begin{equation}
{GM_*\over r^3}{(1+e\cos\varphi)^{3}\over (1+ e^2 + 2e\cos\varphi)^{2}},
\end{equation}
\begin{equation}
E_{12} ={GM_*\over r^3}{3K_z e\sin\varphi (1+e\cos\varphi)
\over N(1+ e^2 + 2e\cos\varphi)}
+{\dot{\bf b}}_1\cdot \dot{\bf b}_2
- {D C_0\over Dt}\nonumber ,\end{equation}
\begin{equation}
E_{21} ={GM_*\over r^3}{3K_z e\sin\varphi (1+e\cos\varphi)
\over N(1+ e^2 + 2e\cos\varphi)}
+{\dot{\bf b}}_1\cdot \dot{\bf b}_2
+ {D C_0\over Dt}\nonumber  \end{equation}
and
\begin{equation}
E_{22} =  {GM_*\over r^3}{{K_z^2(2+4e\cos\varphi +e^2(2-3\sin^2\varphi))}
\over N^2 ( 1+ e^2 + 2e\cos\varphi)} - {\bf b}_2 \ddot{\bf b}_2.
\end{equation}

\end{appendix}
 \newpage
\begin{appendix}\label{AA2}
\title{{\bf APPENDIX 2}}
\section*{Parametric instability for nearly circular streamlines} 
We here consider the generic parametric instability that occurs
for streamlines near to circles for systems with inertia/gravity modes.
 But note that
as indicated for the specific example of Keplerian ellipses
instability may persist for significant 
departures of the streamlines from circles.

\noindent  We start  with the governing 
equations in vector form (see equation (\ref{LBO1}) in section \ref{govsol}).
These are written as:
\begin{equation}
{\bf \ddot y} -2C_0{\bf e_3}\times {\bf \dot y}
= {\bf E}\cdot {\bf  y} ,\label{A2.1}
\end{equation}
where the two dimensional vector $ {\bf  y}^T = (\alpha , \beta)$ and 
the vector ${\bf e_3}\times {\bf \dot y} = (-\dot \beta , \dot \alpha)^T.$
The  two dimensional matrix ${\bf E}$ has elements 
$E_{11}, E_{12}, E_{21}, E_{22}$ as defined  in section \ref{Local}
above.  

\noindent For circular streamlines ${\bf E}={\bf (E)}_c$ and
 $C_0 = C_{0,c}$ are constant, while for small departures from circularity
we can expand to first order in some small parameter $\epsilon.$
When the streamlines are Keplerian ellipses $\epsilon \equiv e.$
We thus write $C_0 = C_{0,c}+\epsilon C_{0,1}$ and
${\bf E}={\bf E}_c + \epsilon{\cal E}.$ 
Here $ C_{0,1}$ and ${\cal E}$ are periodic functions of time
where the period is just the time it takes a fluid element
to orbit around a streamline. Without loss of generality
we may assume that the time average of these 
coefficients over one period is zero.  For the Keplerian ellipse streamlines, 
from the results given in appendix 1, 
$C_{0,c} = \Omega K_z/K,$ and the only non zero element of
${\bf (E)}_c$ is $(E_{22})_c = 3( \Omega  K_z/K)^2.$
Here $\Omega = \sqrt{GM_*/a^3}$ where $a$ is the semi major axis
of the streamline of interest. Similarly, making the equivalence
$\epsilon \equiv e,$
$C_{0,1} = \Omega (K_z/K)^3 \cos \varphi $ while
the elements of ${\cal E}$ are given by
 ${\cal E}_{11} = -\Omega^2 \cos \varphi ,$
${\cal E}_{12} = \Omega^2(K_z/K^3)\sin\varphi(4K_z^2 +2K_r^2),$
${\cal E}_{21} = 2\Omega^2(K_z/K)\sin\varphi  $ and
${\cal E}_{22} = 2\Omega^2(K_z^2/K^4)\cos\varphi(4K_z^2 +K_r^2).$

\noindent Note that we may write to this order $\varphi = \Omega t$
making the coefficients that are proportional to the eccentricity
vary harmonically with time. The period is simply the orbital
period on the streamline. We now write the governing
equations in the form
\begin{equation}
{\bf \ddot y} -2C_{0,c}{\bf e_3}\times {\bf \dot y}
- {\bf (E)}_c \cdot {\bf  y} = 2\epsilon 
C_{0,1}{\bf e_3}\times {\bf \dot y}
 + {\epsilon \cal E}\cdot {\bf  y}\label{A2.2}
\end{equation}
and treat the right hand side using perturbation theory.

\noindent We first remark that when $\epsilon=0,$ there is a simple normal
mode solution appropriate to the time independent case which
may be written in the form

\begin{equation}
{\bf  y} = {\bf  y}_0 \exp(i\sigma t), \end{equation}
where ${\bf  y}_0$ is a constant vector and $\sigma$ is a normal mode
frequency which satisfies

\begin{equation}
-\sigma^2 {\bf y}_0 -2i\sigma C_{0,c}{\bf e_3}\times {\bf  y}_0
-{\bf (E)}_c \cdot {\bf  y}_0  = 0. \label{A2.3}\end{equation}

\noindent In line with the example presented in section \ref{sime},
we expect this mode to correspond to an inertial/gravity mode.
The spectrum is dense and so we may always expect to be able 
to choose $\sigma$ to be close to $\Omega/2$
which is the lowest order commensurability condition for parametric resonance
(for the example of section \ref{sime} the commensurability is exact
for $K_r/K_z = \sqrt{3}$).

\noindent We now move on to consider
the case of $\epsilon$ non zero but small
and set $\sigma = \Omega/2 + \delta \sigma,$
where $\delta \sigma = O(\epsilon \Omega ).$

\noindent We now seek a solution of equation (\ref{A2.2})
in the form 
\begin{equation}
{\bf  y} = A(t){\bf  y}_0 \exp(i\Omega t/2) +cc. +\sum_{n=1}^{\infty}\epsilon^n
 {\bf  y}_n \label{A2.4}\end{equation}
Here $A(t)$ is a  slowly varying amplitude function to  be determined
and  here and below 
$cc$ denotes the addition of the complex conjugate of the first
expression ensuring that as required ${\bf y}$ is real.
The terms $\epsilon^n {\bf  y}_n,$ assumed real   are
additions that are of first and higher  order in $\epsilon.$

\noindent By stating that $A$ is slowly varying we mean that
$|{\dot A}/A| = O(\epsilon \Omega )$
and
$|{\ddot A}/A| = O(\epsilon \Omega )^2 .$

\noindent Given this we substitute
equation (\ref{A2.4}) into equation (\ref{A2.2})
 and  drop expressions 
that are  clearly second or higher order in $\epsilon$ to get an equation for $A.$ This reads
$$\left(i\Omega {\bf y}_0 -2C_{0,c}{\bf e_3}\times {\bf  y}_0\right)
\left(\dot A -i\delta \sigma A\right)\exp(i\Omega t/2) + cc = $$
$$ \hspace{1cm}
\epsilon\left(i\Omega C_{0,1}{\bf e_3}\times {\bf  y}_0
+ {\cal E} {\bf y}_0\right)A\exp(i\Omega t/2) + cc $$
\begin{equation}
\hspace{1cm} -\epsilon \left({\bf \ddot y}_1 -2C_{0,c}{\bf e_3}\times {\bf \dot y}_1
- {\bf (E)}_c \cdot {\bf  y}_1\right)
\end{equation}

\noindent We now multiply by $)\exp(-i\Omega t/2)$ and take a time average
over  many orbital periods, assuming $A$ varies slowly
enough that it can be factored out,  to obtain

$$\left( i\Omega {\bf y}_0 -2C_{0,c}{\bf e_3}\times {\bf  y}_0\right)
\left( \dot A -i\delta \sigma A\right) = \hspace{1cm} $$
$$ \hspace{0.25cm} \langle \epsilon\left(-i\Omega C_{0,1}{\bf e_3}\times {\bf  y}^*_0
+ {\cal E} {\bf y}^*_0\right) \exp(-i\Omega t) \rangle A^*$$

\begin{equation} \hspace{0.5cm} 
-\epsilon \langle\left ({\bf \ddot y}_1 -2C_{0,c}{\bf e_3}\times {\bf \dot y}_1
- {\bf (E)}_c \cdot {\bf  y}_1\right)\exp(-i\Omega t/2)\rangle
\label{A2.5}\end{equation}

\noindent where the angular brackets denote the time average over many orbital periods.
Because $ C_{0,1}$ and ${\cal E}$ vary harmonically with angular
frequency $ \Omega,$ it is
anticipated that the time averages retained above are not zero.

\noindent Setting $A = A_0\exp(\lambda t)$ 
in equation (\ref{A2.5}), $\lambda$ being
the growth rate, taking the scalar product with
${\bf y}_0^*,$  which as can be shown by doing the time 
integration by parts reduces 
the final term involving ${\bf  y}_1$  to be $O(\epsilon^2)$
and finally taking the modulus of the result  gives 
an expression for $\lambda ^2$  correct to $O(\epsilon^2)$ in the form

\begin{equation} \lambda ^2 =
{\left|\langle \epsilon
 {\bf y}^*_0\cdot {\cal E} {\bf y}^*_0 \exp(-i\Omega t) \rangle \right|^2
\over \left|i\Omega |{\bf y}_0|^2  -2C_{0,c} {\bf y}_0^*\cdot
{\bf e_3}\times {\bf  y}_0\right |^2} - (\delta \sigma^2).
\label{A2.6}\end{equation}

\noindent Thus parametric instability occurs ($\lambda$ is real)
at a maximum rate when the commensurability is exact or $\sigma = \Omega/2$
and for fixed
$\epsilon$ the frequency half-width associated with instability corresponds to
this maximum
 growth rate in magnitude. The growth rate itself is $\propto \epsilon.$

\noindent It is easy to evaluate equation (\ref{A2.6}) for the
Keplerian streamline model of section \ref{notsim}. We begin by noting that
${\bf y}_0$ can be found from the solution for the 
local oscillation mode given for the circular
streamline case in section \ref{sime} (noting that $\Omega'=
-3\Omega/(2r)$  for the Keplerian case). From that
${\bf y}_0^T =( 2iK_z\Omega , K\sigma ).$

\noindent It is then straightforward to evaluate $\lambda$ given ${\bf y}, C_{0,c}$
and ${\cal E}$ as specified above. One obtains
\begin{equation} \lambda ^2 = {9 e^2\Omega^2/256}  -(\delta \sigma^2),
\end{equation}
giving a peak growth rate as a function of eccentricity 
given by  $\lambda_{max} = 3e\Omega/16 .$
The instability band width is then given by $\delta \sigma = \lambda_{max}.$

\noindent 

\end{appendix}
\vspace{4mm}
\begin{appendix}\label{A.3}
 \title{{\bf APPENDIX 3}}
\section*{Relation to the Spectrum of the global problem}
We have studied the stability of fluid displacements that are localized
around the vicinity of fluid streamlines and circulate around with the fluid.
When conditions vary periodically around streamlines, parametric
excitation becomes a possibility aided by an underlying
mode spectrum that is dense so that resonance conditions may
always be satisfied. The analysis was done in a local limit
of high wavenumber in directions orthogonal to streamlines.
However, it is possible to relate the local spectrum so found
to that of the global problem. A physical reason for this is that it
is possible to take a local limit where wavenumbers
tend to infinity but mode oscillation frequencies remain finite.
This naturally results in a limit in which the group velocity vanishes
and hence localization becomes possible. We here illustrate
the connection between the local and global problems.

\noindent The equation satisfied by the Lagrangian displacement ${\bmth{\xi}}$
is equation(\ref{1sem}) which after use of equation (\ref{Feq})
and replacement of the convective derivatives through
equations(\ref{ssvconv}) and (\ref{ssvconv2})
yields an operator equation of the form
\begin{equation}
{\cal{L}}({\bmth{\xi}}, \sigma) = 0,
\end{equation}
where we have included the $\sigma$ dependence to emphasize that
we have an eigenvalue problem for $\sigma$ and
that the determined eigenvalues constitute the spectrum of 
the system.
We adopt the inner product defined through (\ref{iprod})
to define the norm $|\bmth{\xi}| = |(\bmth{\xi},\bmth{\xi})|.$

\noindent Then following Papaloizou \& Pringle (1982)
and Terquem \& Papaloizou (1996), we remark that  the spectrum
coincides with the values of $\sigma$
for which the operator  ${\cal{L}}$  has an unbounded inverse.
Further if one can find any $\bmth{\xi}$ for which
$|{\cal{L}}( {\bmth{\xi}}, \sigma )|^2/|\bmth{\xi}|^2$ can be made as small
as we please then ${\cal{L}}$ will have an unbounded inverse
and so $\sigma$ belongs to the spectrum.

\noindent In fact displacements of the form (\ref{locco}) 
which are localized on streamlines over a length scale that 
tends to zero as a  fractional
power of the  wave number parameter $\lambda^{-1},$
which itself  scales as the inverse wavenumber
as that approaches zero, 
 can be used to construct a set of normalized $\bmth{\xi}$ 
which lead to $|{\cal{L}}( {\bmth{\xi}}, \sigma )|^2 \rightarrow 0$
as $\lambda^{-1} \rightarrow 0$
(see Papaloizou \& Pringle 1982; Terquem \& Papaloizou 1996
for other examples).

\noindent In this limit $\bmth{\xi}$ is of the form (\ref{fix}) 
and   is obtained from the eigenvalue problem (\ref{LBO2}), as is $\sigma,$
which then belongs to the spectrum of ${\cal L}.$

\end{appendix}

\end{document}